# A microresonator frequency comb optical clock

Scott B. Papp[1,*], Katja Beha[1], Pascal Del'Haye[1], Franklyn Quinlan[1],
Hansuek Lee[2], Kerry J. Vahala[2], and Scott A. Diddams[1]

[1]Time and Frequency Division 688, National Institute of Standards and Technology, Boulder, CO 80305 USA

[2]T. J. Watson Laboratory of Applied Physics, California Institute of Technology, Pasadena, CA 91125 USA

[*]Correspondence to: scott.papp@nist.gov

**Abstract**: Optical-frequency combs enable measurement precision at the 20th digit, and accuracy entirely commensurate with their reference oscillator. A new direction in experiments is the creation of ultracompact frequency combs by way of nonlinear parametric optics in microresonators. We refer to these as microcombs, and here we report a silicon-chip-based microcomb optical clock that phase-coherently converts an optical-frequency reference to a microwave signal. A low-noise comb spectrum with 25 THz span is generated with a 2 mm diameter silica disk and broadening in nonlinear fiber. This spectrum is stabilized to rubidium frequency references separated by 3.5 THz by controlling two teeth 108 modes apart. The optical clock's output is the electronically countable 33 GHz microcomb line spacing, which features an absolute stability better than the rubidium transitions by the expected factor of 108. Our work demonstrates the comprehensive set of tools needed for interfacing microcombs to state-of-the-art optical clocks.

Optical clocks leverage the narrow, unvarying transitions of atoms to realize exceptionally stable laser frequencies measured at below the $10^{-17}$ level (*1*, *2*). Optical-frequency combs facilitate the measurement and use of these atomic references by providing a dense set of clock-referenced lines that span more than an octave (*3*). By way of their extraordinary measurement precision, frequency combs have enabled advances in diverse fields from spectroscopy of atoms and molecules (*4*, *5*) to astronomy (*6*). Some of the most exciting, but yet unrealized, applications call for operation in environments that are unsuitable for existing comb technology.

A new class of frequency combs has emerged based on optical microresonators (*7*, *8*). Here the comb generation relies on nonlinear parametric oscillation and cascaded four-wave mixing driven by a single CW laser. Such microcombs offer revolutionary advantages including chip-based photonic integration, uniquely large comb-mode spacings in the 10's of GHz range, and monolithic construction with small size and power consumption. Progress in microcomb development has included frequency control of their spectra (*9–12*), characterization of their noise properties (*13–15*), a Rb-stabilized microcomb oscillator (*16*), and demonstration of phase-locked (*13*, *17*, *18*) and modelocked states (*19*, *20*). However, the milestone of all-optical frequency control of a microcomb to an atomic reference, including phase-coherent frequency division to the microwave domain, has not been achieved.

In this paper, we report the achievement of this goal by demonstrating a functional optical clock based on stabilization of a microcomb to atomic Rb transitions. We generate a low-noise, continuously equidistant microcomb spectrum by use of an on-chip silica microresonator. The clock output is the 33 GHz microcomb line spacing, which is electronically measurable, and a traceable, integer partition of the 3.5 THz frequency spacing of the Rb references. Here we explore the basic features of this microcomb clock, and demonstrate that its Allan deviation can reach $10^{-12}$ for 10,000 s of averaging. Our results highlight an architecture for integration of microcombs with other high-performance and chip-scale atomic frequency references (21).

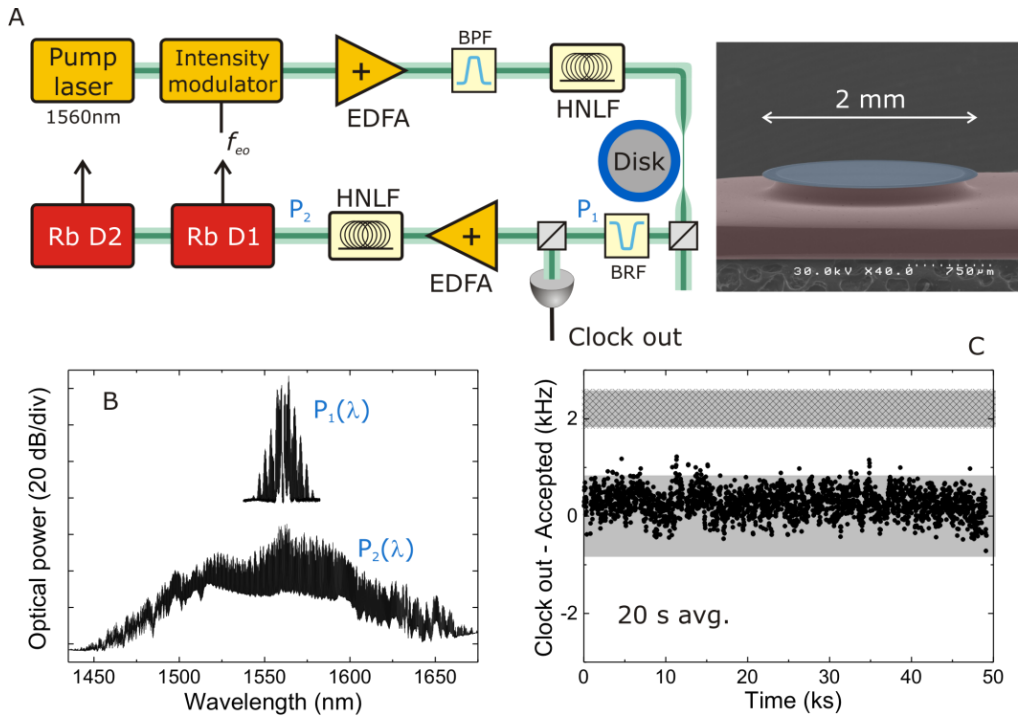

**Figure 1: Microcomb optical clock with Rb atoms.** **(A)** Diagram of our setup. An intensity-modulated pump laser excites a chip-based microresonator (see micrograph at right) to create a 33 GHz spacing comb. The comb is broadened in highly nonlinear fiber (HNLF) following amplification to an average power of 1.4 W. Two lines of the comb 108 apart are stabilized to Rb transitions by control of the pump frequency and the intensity modulation $f_{eo}$. The clock output is obtained via photodetection of the unbroadened spectrum. Not shown are polarization controllers, which are needed before the intensity modulator, the microresonator, and the HNLF; and all the elements of the Rb spectrometers. **(B)** Optical spectrum after a filter to suppress the pump (top) and following spectral broadening (bottom). **(C)** Optical clock output over 12 hours. Each point is the average of twenty 1-s measurements. For comparison, published Rb spectroscopic data on the D2-D1 difference has been subtracted. The solid and hatched gray regions represents data from Ref. (27) and (28).

Figure 1A shows a schematic of our microcomb optical clock. A 2 mm diameter disk resonator with 10-degree wedge side profile provides parametric-comb generation. The resonator, which has an unloaded quality factor of 63 million, is fabricated on a silicon chip using conventional semiconductor fabrication techniques (*22*). Hence, the core of our system is scalable and could be integrated with other on-chip photonic elements, and eventually atomic systems (*21*, *23*). In these experiments we use a tapered fiber for evanescent coupling (*24*, *25*). We excite the disk resonator with light from a CW laser that is intensity-modulated at frequency $f_{eo}$ and amplified to a maximum of 140 mW. The modulation sidebands implement our parametric-seeding technique (*12*), which enables unmatched control of the microcomb line spacing. Here we further demonstrate that parametric seeding enables the complete suppression of undesirable, non-equidistant subcombs. Following generation in the disk resonator, the microcomb output is optically filtered to attenuate the pump laser and modulation sidebands; the resulting spectrum, which spans 20 nm about the 1560 nm pump wavelength, is shown by the top trace in Fig. 1B. By amplifying this comb to 1.4 W and without any dispersion control, we broaden the 20 nm bandwidth to >200 nm. The ~ 2 ps duration optical waveform obtained directly from the microresonator is highly stable and repeatable even for different settings of pump frequency and power, intensity modulation, taper-resonator coupling, and pump polarization. Moreover, the waveform offers sufficient peak intensity for our experiments. The broadened spectrum, which is shown by the bottom trace in Fig. 1B, overlaps with the resonance frequencies of molecules such as HCN, $C_2H_2$, $CO_2$, $CH_4$, and of atomic Rb and K after second-harmonic generation.

To stabilize the microcomb spectrum, we heterodyne its output with telecom-grade semiconductor distributed feedback (DFB) lasers, which are stabilized to well-known Rb transitions. Precise Rb spectroscopic data especially near 780 nm exist, and with attention to systematic effects a stability of $10^{-12}/\sqrt{\tau}$ has been demonstrated for measurement periods $\tau$ up to 2000 s (*26*). Therefore, in this paper we focus only on salient details for controlling the microcomb with these optical references. DFB lasers with up to 60 mW output at 1560 nm and 1590 nm are frequency doubled in PPLN to interrogate the Rb D2 and D1 optical transitions, respectively. The lasers are stabilized to the Rb transitions by use of frequency modulation applied to their drive current. The D2 transition frequencies are known to 7 kHz (*26*), while the D1 levels have been measured with an uncertainty of 179 kHz and 79 kHz in Refs. (*27*) and (*28*). Hence, the D1 uncertainty is the dominant factor in our knowledge of any absolute D2-D1 frequency difference. To operate an optical clock, we stabilize the microcomb's two *independent* degrees of freedom to Rb transitions by leveraging frequency control of its spectrum. The central line of the microcomb is phase-locked to the 1560 nm DFB laser (*10*, *16*). Additionally, the 108th comb line from center, which we obtain via spectral broadening, is phase locked to the 1590 nm DFB laser by tuning the microcomb line spacing through control of $f_{eo}$.

The output of our microcomb optical clock is obtained via photodetection of the 32.9819213 GHz line spacing, which reflects the frequency difference $\Delta_{Rb}$ of the D2- and D1-stabilized lasers divided by 108, and a fixed 660/108 MHz offset for phase stabilization. This offset could assume a range of predetermined values, including zero. The data points in Fig. 1C are a continuous >12 hour long record of the clock output. Here the vertical axis shows the difference between the clock output and $\Delta_{Rb}$ = 3 561 387 470 (180) kHz, which is obtained from Ref. (*27*), and whose uncertainty is shown by the gray band. The hatched band is the result of another measurement of D1 levels from Ref. (*28*). Although we have not systematically analyzed the accuracy of our clock, its output is in reasonable agreement with these previous data. The 271 Hz RMS fluctuation in the clock output is significantly reduced from those of the D2 and D1 reference lasers, owing to the principle of optical frequency division associated with frequency combs (*3*, *29*, *30*). The remainder of this paper presents an investigation of our clock, including deterministic generation of an equidistant microcomb spectrum, demonstration of the precise relationship between the clock output and Rb reference, and analysis of the clock's stability.

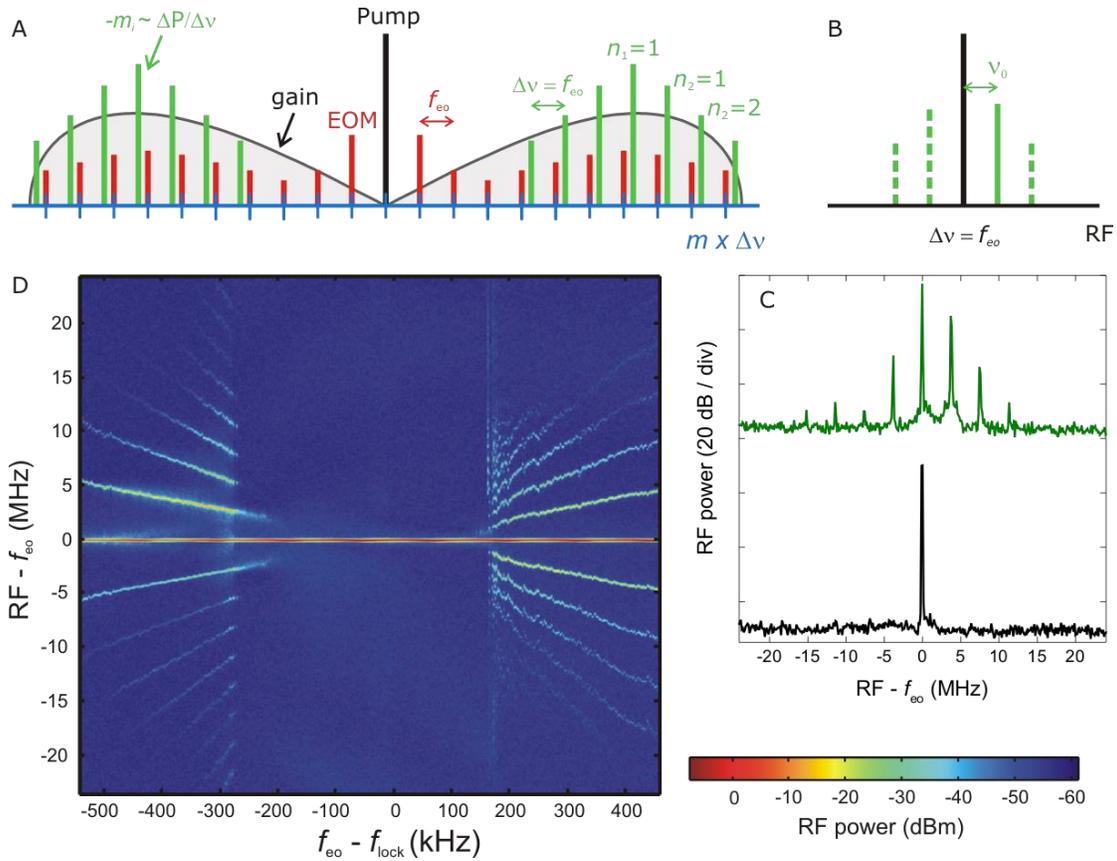

**Figure 2: Deterministic injection locking to create an equidistant microcomb. (A) Model for microcomb generation, including subcomb (green) behavior and parametric seeding (red). (B) Model RF spectrum after photodetection. All the comb lines contribute at frequency $\Delta \nu = f_{eo}$, and the presence of a subcomb is reflected in the sidebands spaced by $\nu_0$. (C) Measured RF spectra indicating a subcomb (top, green) with $\nu_0 \approx 4$ MHz at $f_{eo} - f_{lock} = -387$ kHz and an injection-locked comb (bottom, black) at $f_{eo} = f_{lock}$. (D) Waterfall plot compiled from many traces like those in (C). The false-color bar shows the scaling of RF power.**

The essential aspect of a frequency comb is a uniform spacing of all its modes. However, microcomb spectra are often composed of overlapping, partially nonequidistant subcombs (*31*), which have a fundamentally reduced frequency measurement precision and are unusable for our optical clock experiment. Subcombs arise when parametric oscillation creates initial signal/idler fields in resonator modes $\pm m_i$ apart from the one excited by the pump laser. Their mode spectrum is $v' = v_p + n_1 \Delta P + n_2 \Delta v$, where $\Delta v$ is the fundamental spacing with order $n_2$, $\Delta P \approx m_i \Delta v$ is the primary spacing with subcomb order $n_1$, and $v_p$ represents the pump laser. The subcomb lines have a characteristic frequency offset from that of an equidistant comb given by $v_0 = n_1(\Delta P - m_i \Delta v)$. We use a coherent control technique, parametric seeding (*12*), to deterministically suppress subcombs and favor one with complete equidistance. Figure 2A shows a schematic of the microcomb generation process, including the parametric gain spectrum for our silica resonator, and lines associated both with a subcomb (green) and a seeding comb (red). The outcome of parametric seeding, implemented via intensity modulation of the pump laser at $f_{eo}$, is a comb with frequencies $v_m = v_p + m f_{eo}$ that experiences parametric gain in the microresonator. Moreover, the seeding induces sidebands of the first signal/idler pair that fix $\Delta v$ of all parametrically generated lines at $f_{eo}$. In this paper, we demonstrate for the first time that sufficient amplitude of the seeding comb $v_{\pm m_i}$ modes can injection-lock the subcomb, which completely nullifies its frequency offset $v_0$.

To investigate subcomb injection-locking (Fig. 2), we monitor the amplitudes of subcomb and parametric seeding comb lines while we tune $f_{eo}$, and thus $v_0$. Information about these amplitudes is obtained via photodetection of the entire microcomb spectrum, which yields a signal at frequency $\Delta v$ and its sidebands associated with $v_0$; see the schematic in Fig. 2B. The measurement of RF power versus frequency in Fig. 2C shows $\Delta v$ and $v_0$, including its four-wave mixing harmonics. By recording many such traces for different settings of $f_{eo}$, we explore the transition into and out of injection-locked operation; Fig. 2C presents a false-color "waterfall" plot of these data. Here the horizontal band at zero is the microcomb line spacing, which is always $f_{eo}$, and the other bands are associated with the subcomb offset. As the offset is tuned toward zero via a computer-controlled scan of $f_{eo}$ in 1.7 kHz increments, we observe an abrupt suppression of the subcomb RF components. This represents the point at which the line frequencies of the seeding comb and subcomb are sufficiently close to capture the latter. A linear fit of the first-order offset signals yields the $f_{eo}$ setting ($f_{lock}$) for $v_0 = 0$, which is used to calibrate the horizontal axis of Fig. 2D, and the slope, which corresponds to $m_i$. The injection-locking range is 400 kHz, in which the RF frequencies associated with the subcomb offset are suppressed by >40 dB; see Fig. 2C. Following initiation of injection-locked state, the microcomb's spectrum is equidistant and offset free, and can operate continuously for >24 hours. We verify the equidistance of the central 110 lines of the broadened microcomb spectrum by use of a calibrated reference system; see Refs. (*12*, *32*) for details.

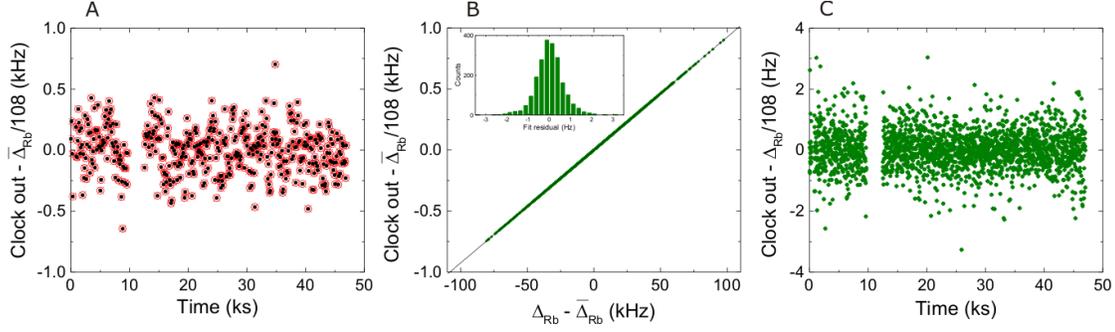

**Figure 3: Sub-division of the 3.5 THz Rb reference. (A)** Clock output (black points) and difference of the Rb-stabilized DFB lasers (red circles) minus the mean of all $\Delta_{Rb}$ measurements. Here 100-sample mean values of the 1-s gate data are displayed. **(B)** Strong correlation of clock output and the THz frequency of the Rb reference. Inset shows a histogram of the fit residuals. **(C)** Real-time-corrected clock output with 0.64 Hz standard deviation. During the gap in measurements at 10 ks, the fiber comb was unlocked.

Our microcomb optical clock is designed to generate an electronically detectable microwave output at a precise *integer* sub-division of the Rb reference. In contrast to earlier work (*16*), the clock output is traceable to atomic structure without regard to the operating parameters or conditions of the microcomb. To test this principle, we simultaneously count the DFB lasers' frequencies and the clock output, and characterize their degree of correlation. The frequencies are recorded in nearly continuous 1-s intervals by use of an auxiliary, self-referenced and repetition-rate stabilized Er:fiber frequency comb (*33*). In Fig. 3A, the black points show the clock output, while the red open points are $\Delta_{Rb}/108$. Since the difference of the DFB lasers is not calibrated, for clarity we subtract its mean value $\overline{\Delta}_{Rb}/108$ from all the points. The overlap of the two data sets suggests their correlation, which we analyze in more detail by plotting them against each other; see Fig. 3B. A linear fit of this correlation plot yields a slope of 108.0002(59), and the horizontal intercept differs from zero by only $-2.4 \pm 1.5$ Hz, compared to the ~3.5 THz frequency of the Rb reference. Knowledge of $\Delta_{Rb}$ via the stabilized fiber comb enables us to compare in real time the clock signal with its optical reference. Figure 3C shows a frequency-counter record of the clock output from the same dataset as in Fig. 1C, but here at each point a correction for noise of the Rb spectrometers is applied. This reduces the scale of clock fluctuations by a factor of ~1000 to the Hz level, and demonstrates that they are overwhelmingly determined by the Rb reference.

We expect the microcomb clock output will closely reproduce the frequency stability of the Rb references. To characterize them, we record the optical heterodyne frequency of the microcomb and the 1590 nm laser, while the microcomb's central line is phase-locked to the 1560 nm laser. For this experiment, the phase-lock to the 1590 nm laser is switched off and a constant parametric seeding frequency, which is synthesized from a hydrogen maser, determines the microcomb line spacing. The open black circles in Fig. 4 show the combined Allan deviation of the Rb references normalized to 33 GHz for six decades of integration time. For short measurement periods the instability decreases as approximately $1/\sqrt{\tau}$. However near 0.1 s the Allan deviation increases and only slowly improves for averaging periods up to 10,000 s. The impact of systematic drifts on the Rb transitions due to, for example, excitation laser power fluctuations, magnetic-field noise, and the Rb vapor cell pressure have been discussed extensively in the literature; eg. Ref. (*26*). Importantly, these effects, rather than any associated with the microcomb, explain the references' stability.

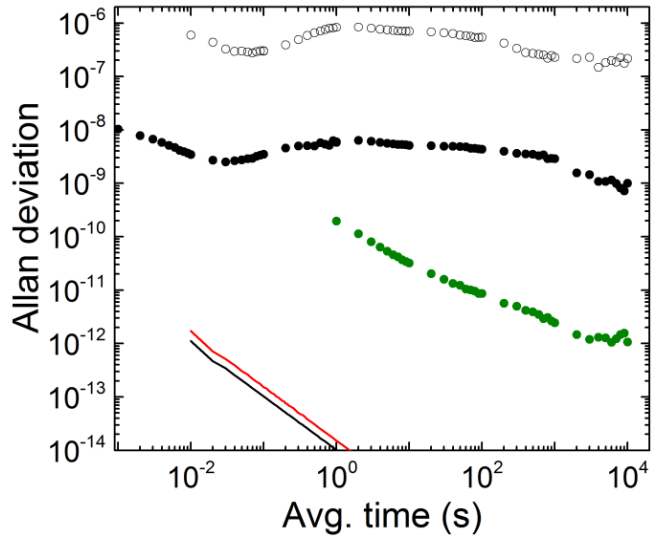

**Figure 4: Microcomb optical clock stability.** Allan deviation of the DFB lasers (open points) and the microcomb clock (closed black points) normalized to a 33 GHz carrier. The green points are the clock stability after post-correcting the Rb reference noise via measurements against an auxiliary comb. The red and black traces represent the in-loop residual noise for stabilization of the micorcomb pump laser and mode 108, respectively.

With microcomb servo control via the Rb D1 reference restored, we analyze the clock output by way of its Allan deviation, which is obtained with respect to a synthesized 33 GHz frequency referenced to a hydrogen maser. Importantly, the clock's stability (filled points in Fig. 4) is improved by a factor of ~100 over that of the DFB lasers, whose noise is distributed among all the lines of the microcomb. Such optical-frequency division is the key metrological feature of any comb. Fluctuations and inaccuracy of the Rb-referenced lasers are naturally reflected in the microcomb clock output, and this explains the slow averaging of the filled points in Fig. 4 beyond ~0.1 s. On the other hand, the in-loop noise of the two servos that stabilize the microcomb, which are shown by the black and red traces in Fig. 4, indicate that our microcomb system as currently configured could support more stable frequency references, and hence produce a 33 GHz output with 1-s fractional stability of $2 \times 10^{-14}$. We demonstrate this potential for improvement by characterizing the clock output, including its real time correction for Rb-reference noise (Fig. 3C). The green points in Fig. 4 show an upper limit for the corrected clock output's Allan deviation, which monotonically decreases with averaging time to $10^{-12}$ at 10,000 s. Accordingly, a microcomb utilizing compact, all-optical Rb frequency references in a controlled environment (*26*) could produce a 33 GHz output with $5 \times 10^{-11}/\sqrt{\tau}$ stability.

In conclusion, we have reported all-optical stabilization of a chip-based microresonator frequency comb to atomic Rb transitions. The electronically detectable microwave output of our optical clock is stable and it accurately subdivides the THz frequency difference of the Rb references. At present the clock output stability is limited entirely by the Rb references. However, the microcomb frequency control architecture demonstrated here is sufficient to support references with orders-of-magnitude higher performance. Future work will address this point, as well as focus on generation of higher peak-power optical waveforms directly from the microcomb. Combined with nonlinear spectral broadening this would enable the subdivision of even larger frequency gaps and ultimately self-referencing (*34*) of a microcomb.

We thank Liz Donley and Andrew Ludlow for thoughtful comments on this manuscript and M. Hirano for supplying the HNLF. This work is supported by the DARPA QuASAR and PULSE programs, NIST, and NASA. It is a contribution of the US government (NIST) and is not subject to copyright in the United States of America.

**References and Notes:**

1. C. W. Chou, D. B. Hume, J. C. J. Koelemeij, D. J. Wineland, T. Rosenband, Frequency Comparison of Two High-Accuracy Al+ Optical Clocks, *Phys. Rev. Lett.* **104**, 070802– (2010).

2. N. Hinkley, J. Sherman, An atomic clock with 10E-18 instability, *ScienceExpress* (2013), doi:10.1126/science.1240420.

3. S. A. Diddams *et al.*, An Optical Clock Based on a Single Trapped 199Hg+ Ion, *Science* **293**, 825–828 (2001).

4. K.-K. Ni *et al.*, A High Phase-Space-Density Gas of Polar Molecules, *Science* **322**, 231–235 (2008).

5. M. J. Thorpe, K. D. Moll, R. J. Jones, B. Safdi, J. Ye, Broadband Cavity Ringdown Spectroscopy for Sensitive and Rapid Molecular Detection, *Science* **311**, 1595–1599 (2006).

6. M. T. Murphy *et al.*, High-precision wavelength calibration of astronomical spectrographs with laser frequency combs, *Mon. Not. R. Astron. Soc.* **380**, 839 (2007).

7. T. J. Kippenberg, R. Holzwarth, S. A. Diddams, Microresonator-Based Optical Frequency Combs, *Science* **332**, 555–559 (2011).

8. P. Del'Haye *et al.*, Optical frequency comb generation from a monolithic microresonator, *Nature* **450**, 1214–1217 (2007).

9. P. Del'Haye, O. Arcizet, A. Schliesser, R. Holzwarth, T. J. Kippenberg, Full Stabilization of a Microresonator-Based Optical Frequency Comb, *Phys. Rev. Lett.* **101**, 53903–53904 (2008).

10. S. B. Papp, P. Del'Haye, S. A. Diddams, Mechanical Control of a Microrod-Resonator Optical Frequency Comb, *Physical Review X* **3**, 31003 (2013).

11. P. Del'Haye, S. B. Papp, S. A. Diddams, Hybrid Electro-Optically Modulated Microcombs, *Phys. Rev. Lett.* **109**, 263901– (2012).

12. S. B. Papp, P. Del'Haye, S. A. Diddams, Parametric seeding of a microresonator optical frequency comb, *Opt. Express* **21**, 17615–17624 (2013).

13. S. B. Papp, S. A. Diddams, Spectral and temporal characterization of a fused-quartz-microresonator optical frequency comb, *Phys. Rev. A* **84**, 053833– (2011).

14. A. A. Savchenkov, E. Rubiola, A. B. Matsko, V. S. Ilchenko, L. Maleki, Phase noise of whispering gallery photonic hyper-parametric microwave oscillators, *Opt. Express* **16**, 4130–4144 (2008).

15. J. Li, H. Lee, T. Chen, K. J. Vahala, Low-Pump-Power, Low-Phase-Noise, and Microwave to Millimeter-Wave Repetition Rate Operation in Microcombs, *Phys. Rev. Lett.* **109**, 233901– (2012).

16. A. A. Savchenkov *et al.*, Stabilization of a Kerr frequency comb oscillator, *Opt. Lett.* **38**, 2636–2639 (2013).

17. F. Ferdous *et al.*, Spectral line-by-line pulse shaping of on-chip microresonator frequency combs, *Nat Photon* **5**, 770–776 (2011).

18. P. Del'Haye, S. B. Papp, S. A. Diddams, Self-Injection Locking and Phase-Locked States in Microresonator-Based Optical Frequency Combs, (2013) (available at http://arxiv.org/abs/1307.4091).

19. T. Herr *et al.*, Mode-locking in an optical microresonator via soliton formation, *arXiv:1211.0733* (2012).

20. K. Saha *et al.*, Modelocking and femtosecond pulse generation in chip-based frequency combs, *Opt. Express* **21**, 1335–1343 (2013).

21. S. Knappe *et al.*, Atomic vapor cells for chip-scale atomic clocks with improved long-term frequency stability, *Opt. Lett.* **30**, 2351–2353 (2005).

22. H. Lee *et al.*, Chemically etched ultrahigh-Q wedge-resonator on a silicon chip, *Nat Photon* **6**, 369–373 (2012).

23. W. Yang *et al.*, Atomic spectroscopy on a chip, *Nat Photon* **1**, 331–335 (2007).

24. M. Cai, O. Painter, K. J. Vahala, Observation of Critical Coupling in a Fiber Taper to a Silica-Microsphere Whispering-Gallery Mode System, *Phys. Rev. Lett.* **85**, 74–77 (2000).

25. S. M. Spillane, T. J. Kippenberg, O. J. Painter, K. J. Vahala, Ideality in a Fiber-Taper-Coupled Microresonator System for Application to Cavity Quantum Electrodynamics, *Physical Review Letters* **91**, 43902 (2003).

26. J. Ye, S. Swartz, P. Jungner, J. L. Hall, Hyperfine structure and absolute frequency of the 87Rb 5P3/2 state, *Opt. Lett.* **21**, 1280–1282 (1996).

27. A. Marian, M. C. Stowe, J. R. Lawall, D. Felinto, J. Ye, United Time-Frequency Spectroscopy for Dynamics and Global Structure, *Science* **306**, 2063–2068 (2004).

28. M. Maric, J. J. McFerran, A. N. Luiten, Frequency-comb spectroscopy of the D1 line in laser-cooled rubidium, *Physical Review A* **77**, 32502 (2008).

29. W. C. Swann, E. Baumann, F. R. Giorgetta, N. R. Newbury, Microwave generation with low residual phase noise from a femtosecond fiber laser with an intracavity electro-optic modulator., *Optics express* **19**, 24387–95 (2011).

30. T. M. Fortier *et al.*, Generation of ultrastable microwaves via optical frequency division, *Nat. Photon.* **5**, 425 (2011).

31. T. Herr *et al.*, Universal formation dynamics and noise of Kerr-frequency combs in microresonators, *Nat Photon* **6**, 480–487 (2012).

32. See Supporting Online Material.

33. F. Quinlan *et al.*, Ultralow phase noise microwave generation with an Er:fiber-based optical frequency divider, *Opt. Lett.* **36**, 3260–3262 (2011).

34. D. J. Jones *et al.*, Carrier-Envelope Phase Control of Femtosecond Mode-Locked Lasers and Direct Optical Frequency Synthesis, *Science* **288**, 635–639 (2000).